%% file: msv7.tex
\documentclass[12pt,preprint,useAMS,usenatbib,usegraphicx,english]{aastex}

\usepackage[T1]{fontenc}
\usepackage[utf8]{inputenc}
\usepackage{babel}

\shortauthors{CASTRO & GIZIS}
\shorttitle{WISE J231921.92+764544.4}
\begin{document}

\title{DISCOVERY OF AN L4$\beta$ CANDIDATE MEMBER OF ARGUS IN THE PLANETARY MASS REGIME: WISE J231921.92+764544.4}

\author{Philip J. Castro and John E. Gizis}
\affil{Department of Physics and Astronomy, University of Delaware, Newark, DE 19716, USA; philip.j.castro@gmail.com, gizis@udel.edu}

\begin{abstract}
We present the discovery of a young L dwarf, WISE J231921.92+764544.4, identified by comparing the 
Wide-field Infrared Survey Explorer (WISE) All-Sky Catalog to the Two Micron All Sky Survey (2MASS). A medium-resolution optical 
spectrum provides a spectral type of L4$\beta$, with a photometric distance estimate of $26.1\pm4.4$ pc.
The red WISE $W1-W2$ color provides additional evidence of youth, while the 2MASS $J-K_{\rm s}$ color does not.
WISE J231921.92+764544.4 is a candidate member of the young moving group Argus, with the
space motion and position of WISE J231921.92+764544.4 giving a probability of 79\% membership in Argus and a probability
of 21\% as a field object, based on BANYAN II. WISE J231921.92+764544.4 has a mass of 12.1$\pm$0.4 M$_{\rm Jup}$ 
based on membership in Argus, within the planetary mass regime.
\end{abstract}

\keywords{
brown dwarfs -
infrared: stars -
proper motions -
stars: distances -
stars: individual (WISE J231921.92+764544.4) -
stars: late-type
}

\section{INTRODUCTION}
Brown dwarfs are objects that lie between the mass range of stars and 
planets (13 M$_{\rm Jup}$ $\lesssim$ M $\lesssim$ 75 M$_{\rm Jup}$), 
they have insufficient mass to sustain hydrogen fusion but enough mass to burn at least deuterium.
Brown dwarfs are brightest when they are born and continously dim and cool thereafter \citep{Basri2000},
and, as a consequence, evolve through the MLT spectral sequence.
An early L dwarf could 
be an old very low mass star, a young brown dwarf, or an even younger planetary 
mass object \citep{Cruzetal2009,Kirkpatrick2005}.
Brown dwarfs contract as they age, with the radius of young brown dwarfs being as much as three times 
greater than their eventual equilibrium state \citep{McGovernetal2004,Burrowsetal2001}.
With g $\propto$ M/R$^{2}$, brown dwarfs evolve from low to high surface gravity as they 
age \citep{McGovernetal2004,Kirkpatrick2005}.

G 196-3B was discovered by \citet{Reboloetal1998} as a companion to a young M2.5 dwarf (G 196-3A, $\sim100$ Myr) 
by direct imaging. The low-resolution optical spectrum indicated a candidate L dwarf and showed 
weak Na I, Rb I, and Cs I, interpreted as an indication of low surface gravity. Follow-up optical observations by 
\citet{Kirkpatricketal2001} showed it to be an L2 dwarf.
Observations of G 196-3B by \citet{McGovernetal2004} in the near-infrared ($J$ band) are supportive of 
low surface gravity, showing the presence of TiO and VO, weak K I lines, and weak FeH absorption.
\citet{McGovernetal2004} showed that TiO, VO, K I, Na I, Cs I, Rb I, CaH, and FeH are gravity sensitive features
in brown dwarfs by comparing these features in late-type giants and in old field dwarfs using low-resolution
$J$ band and optical spectra.
2MASS J01415823-4633574 was discovered by \citet{Kirkpatricketal2006}, classified as L0 pec, and is another 
benchmark object that set the framework for understanding young L dwarfs in the field. Its optical spectrum 
showed very strong bands of VO and weak absorptions by TiO, K I, and Na I, with low-gravity being an explanation 
for these unique spectral signatures. 
The near-infrared spectrum showed a triangular-shaped $H$ band, where this feature was reported in the spectra 
of young brown dwarf candidates in the Orion Nebula Cluster \citep{Lucasetal2001}, providing further evidence 
of youth. The spectrum was also much redder than spectra of normal late-M/early-L dwarfs, its 2MASS $J$-$K_{\rm s}$ 
color being significantly redder than the average color for normal field L0 dwarfs.
The spectrum showed strong VO and weak Na I, K I, and FeH, analogous to the peculiarities seen in the 
optical spectrum.
Numerous other young L dwarfs have been identified in the field \citep{Cruzetal2007,Reidetal2008,Kirkpatricketal2008},
culminating with a preliminary spectral sequence for low-gravity L dwarfs by \citet{Cruzetal2009}.
This sequence includes very low-gravity ($\gamma$) and intermediate-gravity ($\beta$) L dwarfs spanning
L0 to L5, with the greek suffix appending the spectral type to indicate gravity type as suggested 
by \citet{Kirkpatrick2005}. The low-gravity ($\gamma$) are estimated to be closer to $\approx10$ Myr, and the intermediate 
gravity ($\beta$) are more likely $\approx100$ Myr \citep{Cruzetal2009}.
Young brown dwarfs play an important role in that they are analogs to giant exoplanets.
Brown dwarfs and giant exoplanets share overlapping temperature regimes, condensate clouds in their atmospheres,
and have similar photometric and spectroscopic characteristics.
Current technology only allows a handful of planetary systems to be directly studied. With young brown dwarfs
being numerous, bright, and isolated in the field, they serve as excellent candidates for extensive studies
that are not feasible with exoplanets.
Additionally, young brown dwarfs associated with young moving groups provide a more precise age estimate 
than other young brown dwarfs in the field and will therefore play a role in helping to constrain 
substellar evolutionary models \citep{Fahertyetal2013b}.

The Wide-field Infrared Survey Explorer (WISE) all-sky data release (occurred on March 14, 2012)
covers the entire sky in four bands centered at
wavelengths 3.4$\mu$m ($W1$), 4.6$\mu$m ($W2$), 12$\mu$m ($W3$), and 22$\mu$m ($W4$), and achieves 5$\sigma$ detections
for point sources \citep{Wrightetal2010}.
The Two Micron All Sky Survey (2MASS) is a near-infrared survey performed from 1997 to 2001 covering virtually the
entire sky at wavelengths 1.25$\mu$m ($J$), 1.65$\mu$m ($H$), and 2.16$\mu$m ($K_{\rm s}$),
providing a 10$\sigma$ point-source detection level of better than 15.8, 15.1, and 14.3 mag,
respectively \citep{Skrutskieetal2006}.
These two all-sky surveys, with a difference in epochs of $\sim10$ yr, and photometry in the near and mid-infrared,
provide an ideal setup to search for low-mass stars and brown dwarfs via their apparent motion.
Multi-epoch searches using 2MASS and WISE have yielded numerous
discoveries \citep[e.g.,][]{CastroGizis2012,Gizisetal2012,Luhmanetal2012,Luhman2013,Castroetal2013}.

This manuscript presents the discovery of a young brown dwarf, with a mass estimate in the
planetary mass regime based on membership in Argus.
In Section 2 we discuss the discovery and observations of WISE J231921.92+764544.4,
Section 3 the analysis, which includes the optical spectrum, photometric colors, and the distance and physical properties, 
and lastly Section 4 with the conclusions and future work.

\section{DISCOVERY AND OBSERVATIONS}
We have performed a search for objects that have moderate apparent motion between 2MASS and the
WISE all-sky data release\footnote{http://wise2.ipac.caltech.edu/docs/release/allsky/expsup/}, complimentary 
to previous searches \citep{Gizisetal2011b,Gizisetal2011a,CastroGizis2012,Gizisetal2012,Castroetal2013}.
Since WISE is already matched to 2MASS within 3$\arcsec$, and we are searching for objects with moderate proper motion,
we can use the 2MASS information within the WISE catalog in addition to the WISE parameters to help constrain the search.
One of our searches required a detection in $W1$, $W2$, and $W3$, an extended source flag of zero, 
`PH\_QUAL'\footnote{The `PH\_QUAL' flag is a measure of the
photometric quality in each band, with flags A, B, C, U, X, and Z.
A to C represent detections with a decreasing signal
to noise. For more details refer to the Explanatory Supplement to the WISE
All-Sky Data Release Products, http://wise2.ipac.caltech.edu/docs/release/allsky/expsup/sec2\_2a.html\#ph\_qual}
of `AAA' for $W1$, $W2$, and $W3$, color constraints for WISE and 2MASS of
$0.4\le W1$-$W2\le0.8$, $0.8\le W2$-$W3\le1.2$, $W1\le13.0$, $1.55\le J$-$K_{\rm s}\le1.95$,
and the distance between a WISE source and a 2MASS source of $\ge1.0\arcsec$.
WISE sources meeting our criteria were examined visually using 2MASS and WISE
finder charts\footnote{Finder charts at IRSA can be found at http://irsa.ipac.caltech.edu/} in order to
look for apparent motion of a source between the two surveys.
This search criteria yielded six results, five were spurious, while one of them 
was real, WISE J231921.91+764544.3. Although the original search that discovered WISE J231921.91+764544.3 was done using
the WISE all-sky data release, we present our results using the most recent reprocessed
AllWISE data release\footnote{http://wise2.ipac.caltech.edu/docs/release/allwise/expsup/} for WISE J231921.91+764544.3,
using the AllWISE designation of WISE J231921.92+764544.4 (W2319+7645, hereafter).

W2319+7645 is found a distance of 2.0$\arcsec$ to the northeast of 2MASS source 2MASS J23192137+7645437.
The WISE source shows colors that are red, $W1-W2=0.48\pm0.12$, consistent with that of 
a late L dwarf \citep{Kirkpatricketal2011}; the 2MASS source has red colors, $J-K_{\rm s}=1.71\pm0.08$,
that are consistent with an L dwarf \citep{Kirkpatricketal2000}.
Figure 1 shows the 2MASS $K_{\rm s}$ and WISE $W1$, $W2$, $W3$, and $W4$ images centered on W2319+7645,
in all of the images the red circle shows the location of W2319+7645 based on the WISE epoch
and in the $K_{\rm s}$ image the green circle shows the location of W2319+7645 based on the 2MASS epoch.

\begin{figure}
\caption{
2MASS $K_{\rm s}$ and WISE $W1$, $W2$, $W3$, and $W4$ images showing the location of W2319+7645.
The red circles show the location of W2319+7645 based on the WISE epoch and the
green circle shows the location of W2319+7645 based on the 2MASS epoch.
Each image is 4$\arcmin$ x 4$\arcmin$. North is up and east is to the left.
}
\begin{center}
\includegraphics[width=6.5in]{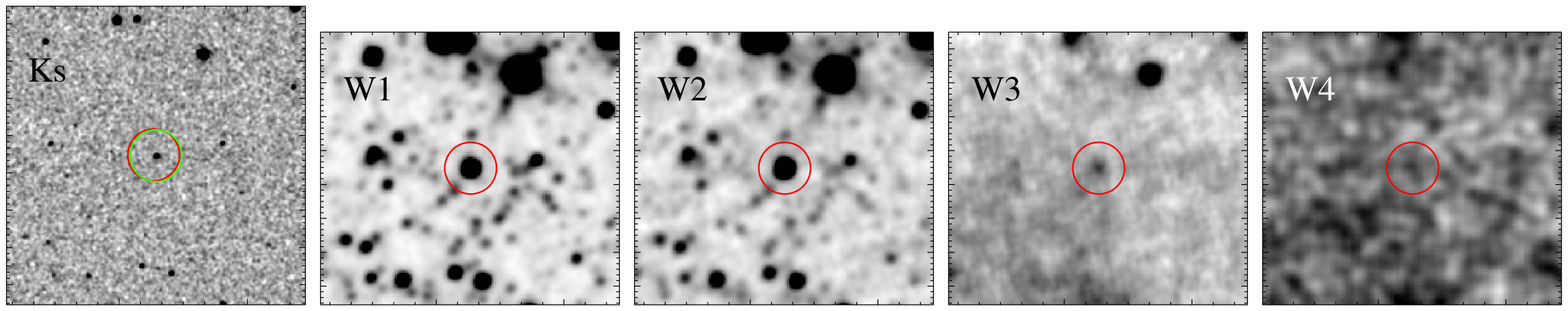}
\end{center}
\end{figure}

W2319+7645 was observed using the Gemini-North telescope. The Gemini-North observations (Gemini program GN-2012B-Q-105)
were on UT Date 11 September 2012 with the GMOS spectrograph \citep{Hooketal2004} using grating R831, and consisted
of four 600 second exposures. The wavelength coverage was 6340 to 8460\AA~with a resolution of $\sim2$\AA. Conditions
were non-photometric. All spectra were processed using standard IRAF tasks.

\section{ANALYSIS}

\subsection{Optical Spectrum}
Figure 2 shows the Gemini GMOS-N spectrum of W2319+7645 (black) compared to the L0 through L5 optical standards (red).
The standards are as follows:
2MASP J0345432+254023 \citep[L0;][]{Kirkpatricketal1999},
2MASSW J1439284+192915 \citep[L1;][]{Kirkpatricketal1999},
Kelu-1 \citep[L2;][]{Kirkpatricketal1999},
DENIS-P J1058.7-1548 \citep[L3;][]{Delfosseetal1997},
2MASSW J1155009+230706 \citep[L4;][]{Kirkpatricketal1999},
and DENIS-P J1228.2-1547 \citep[L5;][]{Delfosseetal1997}.
The TiO head at 7053 \AA\ has a maximum at about spectral type M8, and disappears at 
about L2 \citep{Kirkpatricketal1999}.
The lack of TiO absorption at 7053 \AA\ and red L dwarf J-K$_{\rm s}$ colors indicates that W2319+7645
is an L dwarf later than L1.
We performed a by-eye comparison of the overall continuum of W2319+7645 
to the L0-L5 standards \citep{Kirkpatricketal1999}.
W2319+7645 best fits the L2 standard and is a poor fit to the other L dwarf standards.
VO has a broad trough at about 7334-7534 \AA\ and about 7851-7973 \AA, has a maximum at about spectral type M9,
and has disappeared at $\sim$L4 \citep{Kirkpatricketal1999}.
The L2 standard shows VO present at 7334-7534 \AA\ and 7851-7973 \AA, however,
W2319+7645 does not have any evidence of VO in these wavelength regions.
The lack of VO in W2319+7645 indicates it is an $\sim$L4 or later, therefore, W2319+7645 is not consistent 
with a spectral type of L2.
With W2319+7645 not being consistent with
any of the normal L dwarf standards we investigate the potential of it being a low-gravity L dwarf.

\begin{figure}
\caption{
Gemini GMOS-N spectrum of W2319+7645 (black) compared to the L0 through L5 optical standards (red).
The standards are as follows:
2MASP J0345432+254023 \citep[L0;][]{Kirkpatricketal1999},
2MASSW J1439284+192915 \citep[L1;][]{Kirkpatricketal1999},
Kelu-1 \citep[L2;][]{Kirkpatricketal1999},
DENIS-P J1058.7-1548 \citep[L3;][]{Delfosseetal1997},
2MASSW J1155009+230706 \citep[L4;][]{Kirkpatricketal1999},
and DENIS-P J1228.2-1547 \citep[L5;][]{Delfosseetal1997}.
All spectra are normalized to the mean of the window 8240-8260 \AA.
The y-axis is logarithmic and the spectra are offset vertically by a multiplicative constant.
Several spectral features are labeled. The dashed lines delineate the relatively 
gravity-insensitive region (8000-8400 \AA) \citep{Cruzetal2009}.
}
\begin{center}
\includegraphics[width=5.0in]{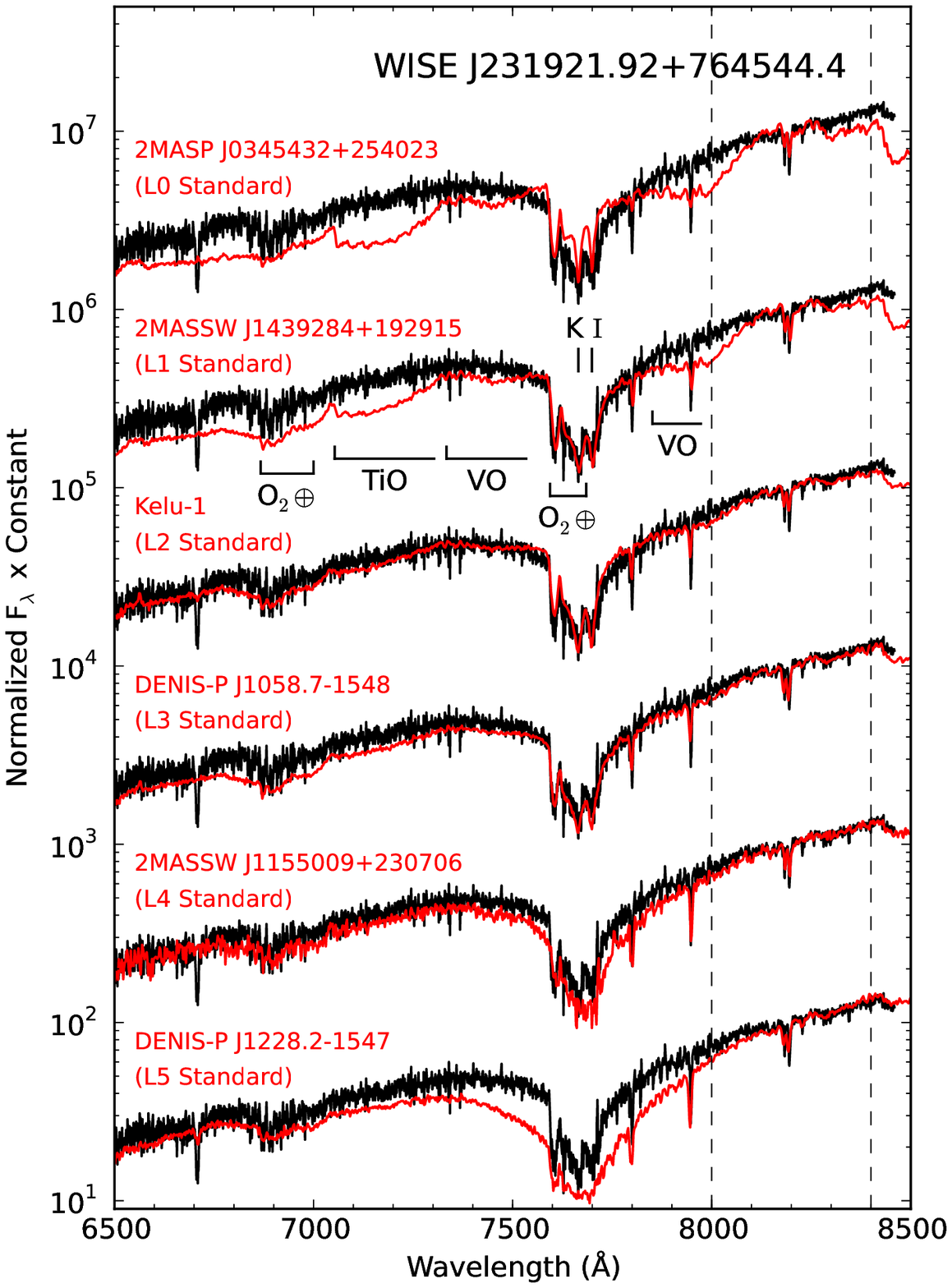}
\end{center}
\end{figure}

As suggested by \citet{Cruzetal2009}, we use a by-eye analysis to compare W2319+7645 to L dwarf standards and 
low-gravity L dwarfs (from \citet{Cruzetal2009}), along with spectral indices, in order to investigate low-gravity 
and consequently determine spectral type.
We compare the overall continuum in the relatively gravity-insensitive region (8000-8400 \AA) \citep{Cruzetal2009} of W2319+7645 to the
L0-L5 standards. A comparison using the gravity-insensitive region will aid in constraining the low-gravity spectral 
subtype of W2319+7645. W2319+7645 best fits the the gravity-insensitive region of the L4 standard.
W2319+7645 has too much flux for the K I doublet and the surrounding region compared to the L4 standard.
We compare the overall continuum of W2319+7645 to the low-gravity L dwarf subtypes of \citet{Cruzetal2009}; with 
the exception of the L3$\beta$ subtype whose spectrum was not available for comparison.
Of the low-gravity L dwarfs compared to, W2319+7645 is a fit to the L4$\beta$, L4$\gamma$, L5$\beta$, and L5$\gamma$.
With a best fit to the gravity-insensitive region of the L0-L5 standards being the L4 standard, 
and W2319+7645 being a fit to the low-gravity L4 dwarfs, we place the focus on a
spectral type of L4$\beta$ and L4$\gamma$.
Figure 3 shows the Gemini GMOS-N spectrum of W2319+7645 (black) compared to
the L4 standard 2MASSW J1155009+230706 (red) \citep{Kirkpatricketal1999},
the intermediate-gravity L4$\beta$ 2MASS J00332386-1521309 (blue),
and the very low-gravity L4$\gamma$ 2MASS J05012406-0010452 (green) \citep{Cruzetal2009}.
Weaker absorption due to the K I doublet is the most prominent feature of the low-gravity L dwarfs, with the later-type (L3-L5)
low-gravity L dwarfs showing more flux than normal L dwarfs near the K I doublet (7300-8000 \AA) \citep{Cruzetal2009}.
W2319+7645 shows weaker absorption of the K I doublet and the surrounding region compared to the L4 standard.
Of the low-gravity L4 dwarfs, W2319+7645 best fits the overall continuum of
the L4$\gamma$. However, the K I doublet of the L4$\gamma$ is much sharper than that of W2319+7645.
W2319+7645 has a slight lack of flux in the region surrounding the K I doublet compared to the L4$\beta$, in 
the region most affected by gravity (7300-8000 \AA) \citep{Cruzetal2009},
as well as for wavelengths shortward of 7300 \AA.
The Na I doublet at 8183 and 8195 \AA\ is either not present or very weak in low-gravity L dwarfs \citep{Cruzetal2009}.
The Na I doublet of the L4 standard and the L4$\beta$
are similar, with W2319+7645 matching them quite well, while the L4$\gamma$ is very weak in comparison; see the zoomed 
in view of the Na I doublet to the right of each spectrum. 
The Na I doublet is a distinctive feature distinguishing the normal-gravity L4 standard and the intermediate-gravity L4$\beta$ from the
very low-gravity L4$\gamma$. With the Na I doublet of W2319+7645 being a poor match to the L4$\gamma$, 
we conclude a by-eye spectral type of L4$\beta$.

\begin{figure}
\caption{
Gemini GMOS-N spectrum of W2319+7645 (black) compared to
the L4 standard 2MASSW J1155009+230706 (red) \citep{Kirkpatricketal1999},
the intermediate-gravity L4$\beta$ 2MASS J00332386-1521309 (blue),
and the very low-gravity L4$\gamma$ 2MASS J05012406-0010452 (green) \citep{Cruzetal2009}.
A zoomed in view of the Na I doublet is shown to the right of each spectrum.
All spectra are normalized to the mean of the window 8240-8260 \AA.
The y-axis is logarithmic and the spectra are offset vertically by a multiplicative constant.
The spectra are binned to the resolution of the L4 standard, which has the lowest resolution spectrum at 1.9 \AA.
Prominent spectral features are labeled. The dashed lines delineate the region most affected by gravity (7300-8000 \AA)
and the relatively gravity-insensitive region (8000-8400 \AA) from \citet{Cruzetal2009}.
}
\begin{center}
\includegraphics[width=6.5in]{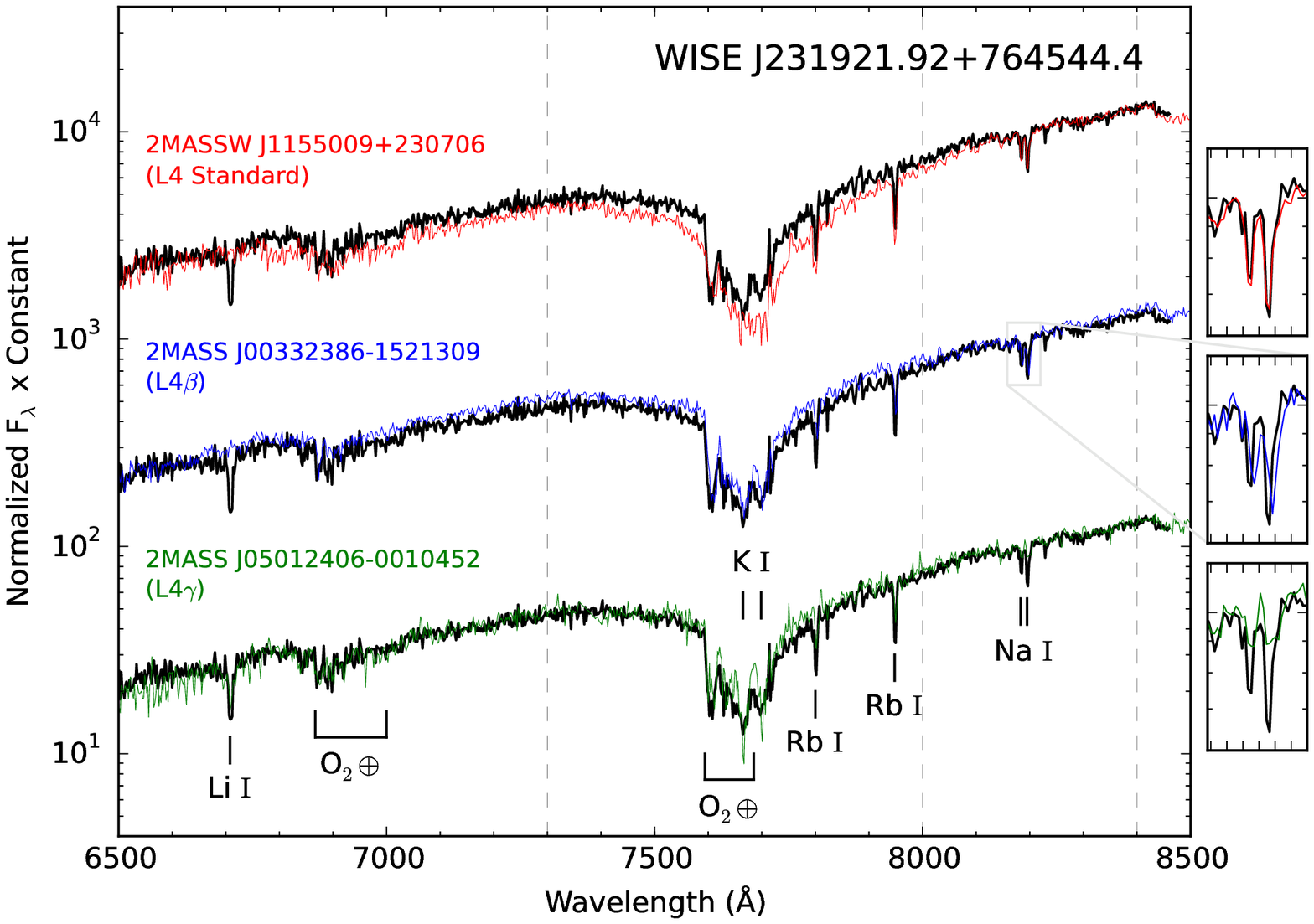}
\end{center}
\end{figure}

Figure 4 shows the spectral indices for W2319+7645 (grey star), 
the L4 standard 2MASSW J1155009+230706 (red circle), the L4$\beta$ 2MASS J00332386-1521309 (blue square),
and the L4$\gamma$ 2MASS J05012406-0010452 (green diamond), based on spectral indices 
defined by \citet{Kirkpatricketal1999} and \citet{Cruzetal2009}.
These spectral indices were shown by \citet{Cruzetal2009} to 
be gravity sensitive. Error bars for the spectral indices were determined following \citet{Cruzetal2009}.
The K-a, K-b, and Rb-b indices strongly indicate that W2319+7645 is low-gravity, while Rb-a 
supports low-gravity. The indices overall are consistent with a spectral
type of L4$\beta$ or L4$\gamma$. Based on our by-eye analysis of the spectrum of W2319+7645 and its spectral indices, 
we find a spectral type for W2319+7645 of L4$\beta$.

\begin{figure}
\caption{
Spectral indices of W2319+7645 (grey star), the L4 standard (red circle), the L4$\beta$ (blue square),
and the L4$\gamma$ (green diamond), using the spectral indices defined by \citet{Kirkpatricketal1999,Cruzetal2009}.
For each index the data points are offset horizontally for clarity.
The spectral indices overall show W2319+7645 to be low-gravity, and are most consistent with an L4$\beta$ or L4$\gamma$.
}
\begin{center}
\includegraphics[width=6.5in]{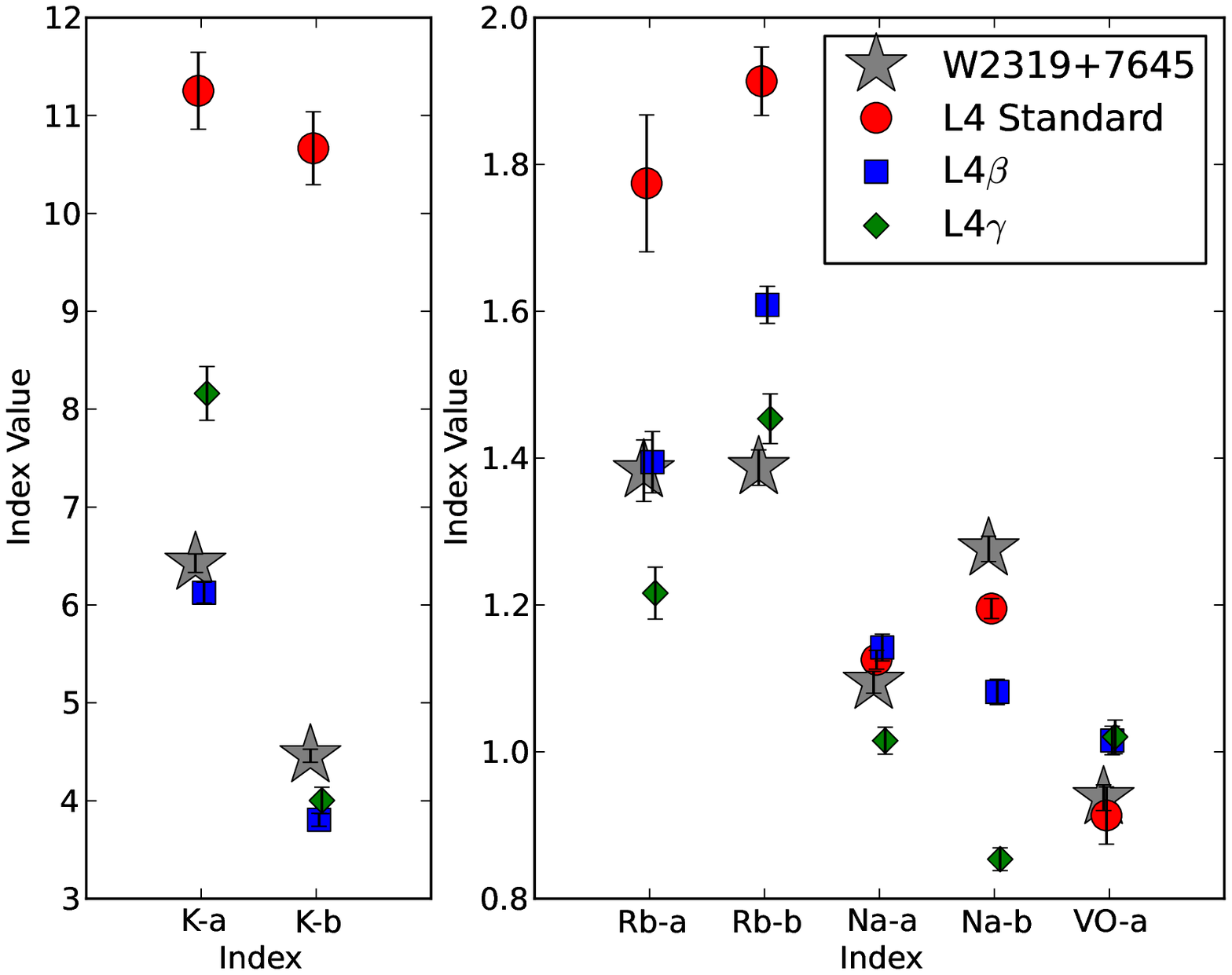}
\end{center}
\end{figure}

The unresolved absorption doublet at 6708 \AA, Li I, is clearly present in the spectrum of W2319+7645.
W2319+7645 passes the lithium test \citep{Reboloetal1992}.
At about 1800 K (L4 spectral subtype), 50\% lithium depletion occurs at about 1 Gyr and a mass of 
about 0.06 M$_{\odot}$ \citep{Kirkpatrick2005}. 
The lithium absorption of W2319+7645 indicates substellarity and youth (< 1 Gyr).
The lack of Li I absorption in the L4$\beta$ and the presence of Li I absorption in the L4$\gamma$ cannot be used as an indicator of youth.
Li I absorption is expected to be weaker for lower gravity
early to mid L dwarfs \citep{Kirkpatricketal2008}, in contradiction to the situation shown for the L4$\beta$ and L4$\gamma$.
Li I absorption is still not fully understood in young brown dwarfs \citep{Kirkpatricketal2008, Cruzetal2009}.

\subsection{Photometric Colors}
Low-gravity L dwarfs have been shown to have red near-infrared colors ($J$-$K_{\rm s}$),
and that they tend to fall at the red end of the distribution within their spectral class \citep{Cruzetal2009}.
\citet{Fahertyetal2013} demonstrated that L$\gamma$ dwarfs tend to lie at the red end of the distribution within
their spectral class for near-infrared colors ($J$-$K_{\rm s}$) and analogously for mid-infrared colors ($W1$-$W2$).
Figure 5 shows 2MASS $J$-$K_{\rm s}$ (top) and WISE $W1$-$W2$ (bottom) colors as a function of spectral type for
W2319+7645 (gray star), normal L dwarfs (red circle), L$\beta$ dwarfs (blue squares), 
and L$\gamma$ dwarfs (green diamonds) for spectral types L0 to L5. 
The normal L dwarfs average values of $J$-$K_{\rm s}$ and $W1$-$W2$, and their standard deviations,
are from \citet{Fahertyetal2013}; we note that the $W1$-$W2$ is All-sky WISE data.
The L$\beta$ and L$\gamma$ dwarfs, and their $J$-$K_{\rm s}$ colors are from \citet{Cruzetal2009},
while the $W1$-$W2$ colors were retrieved from AllWISE.
Just as the L$\beta$ dwarfs tend to be at the red end of the distribution of their spectral types in $J$-$K_{\rm s}$, 
this appears to be the case as well for $W1$-$W2$ colors. We note there are two L$\beta$ dwarfs, one L1$\beta$ and one L4$\beta$,
that lie at the average color of the normal L dwarfs within their spectral class and are not red in $W1$-$W2$ colors.
The $J$-$K_{\rm s}$ color of W2319+7645 is consistent with normal L4 dwarfs and does not provide 
additional evidence of youth. 
This is similar to one of the L1$\beta$ dwarfs that has a $J$-$K_{\rm s}$ color
consistent with normal L1 dwarfs.
The $W1$-$W2$ color of W2319+7645 is at the red end of the distribution for normal L4 dwarfs; 
albeit with large error bars. While an L dwarf having red colors alone does not imply youth, unusually 
thick clouds can also result in red colors \citep{Cruzetal2009}, the red $W1$-$W2$ color of W2319+7645 
combined with the optical spectral type of L4$\beta$ provides additional evidence of youth.

\begin{figure}
\caption{
2MASS $J$-$K_{\rm s}$ (top) and WISE $W1$-$W2$ (bottom) colors as a function of spectral type for L0 to L5 dwarfs.
W2319+7645 (gray star) is compared to normal L dwarfs (red circle) \citep{Fahertyetal2013}, L$\beta$ dwarfs (blue squares), 
and L$\gamma$ dwarfs (green diamonds) \citep{Cruzetal2009}. For each spectral type the data points (except for the normal L dwarfs) 
are offset horizontally for clarity.
}
\begin{center}
\includegraphics[width=6.75in]{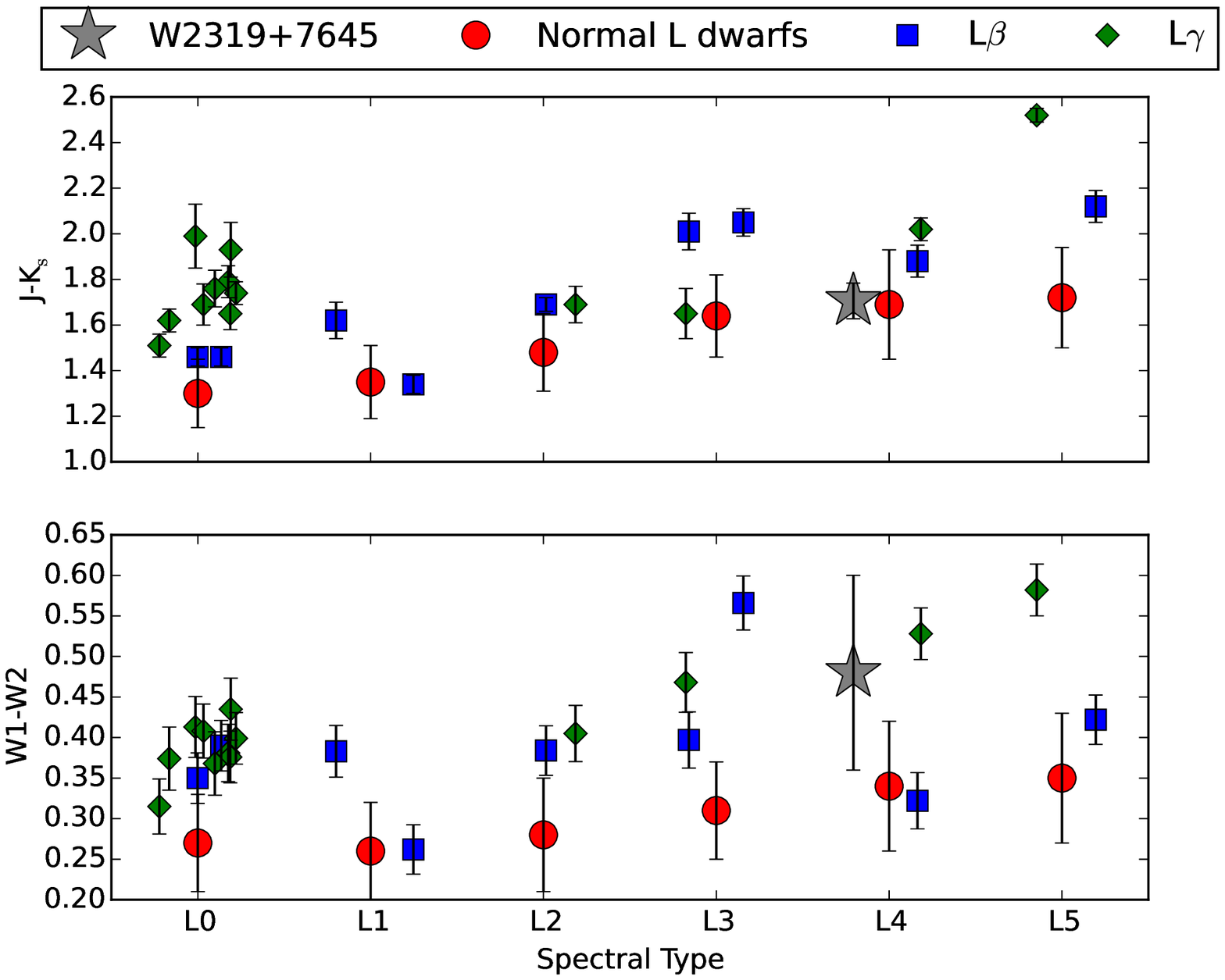}
\end{center}
\end{figure}

\subsection{Distance and Physical Properties}
We provide a crude distance estimate by using the spectral-type-absolute-magnitude relationships of \citet{Looperetal2008a}
for 2MASS photometry and the spectral-type-absolute-magnitude relationships of \citet{DupuyLiu2012} for 2MASS 
and WISE photometry. We find a distance of $26.8\pm3.7$ pc from 2MASS $J$ photometry, 
$27.7\pm3.8$ pc from 2MASS $H$ photometry, and $26.9\pm4.1$ pc from 2MASS $K_{\rm s}$ photometry using the relations 
from \citet{Looperetal2008a}, $27.6\pm5.0$ pc from 2MASS $J$ photometry, $28.2\pm5.2$ pc from 2MASS $H$ photometry, 
$25.4\pm5.0$ pc from 2MASS $K_{\rm s}$ photometry, $24.1\pm4.7$ pc from WISE $W1$ photometry, and
$22.5\pm3.8$ pc from WISE $W2$ photometry using the relations from \citet{DupuyLiu2012}. The uncertainty in the distance 
estimates comes from the uncertainty in the photometry and the RMS from the spectral-type-absolute-magnitude relationships. 
The mean of these estimates provides a distance of $26.1\pm4.4$ pc, assuming no binarity. Trigonometric parallax measurements 
are needed for a reliable distance estimate.

Based on a direct comparison of the 2MASS and WISE positions, W2319+7645 has a proper motion of
$\mu_{\alpha}$cos($\delta$)$=0.19\pm0\farcs01$ yr$^{-1}$
and $\mu_{\delta}=0.07\pm0\farcs01$ yr$^{-1}$, with total apparent motion $0.20\pm0\farcs01$ yr$^{-1}$. 
Based on the estimated distance and apparent motion, W2319+7645 has a tangential velocity of $25\pm5$ km s$^{-1}$.
A radial velocity of $10\pm5$ km s$^{-1}$ was measured for W2319+7645 using the narrow atomic lines in the Gemini spectrum,
giving a total space velocity of $27\pm7$ km s$^{-1}$.
\citet{Fahertyetal2009} gives a median tangential velocity for L4 dwarfs of 25 km s$^{-1}$ with a dispersion of 20 km s$^{-1}$,
while the median tangential velocity for low surface gravity dwarfs is 18 km s$^{-1}$ with a 
dispersion of 15 km s$^{-1}$. W2319+7645 is consistent with both of these populations.
Using the position, motion, and distance of W2319+7645, from \citet{Gagneetal2014} we 
determine\footnote{Gagn\'e, Jonathan; Astrolib (2014): IDL routines to compute XYZ and UVW coordinates. figshare. \newline
http://dx.doi.org/10.6084/m9.figshare.899753 \newline
Retrieved 21:06, Aug 24, 2014 (GMT)} 
a heliocentric galactic position of
(X, Y, Z)=(-12$\pm$2, 22$\pm$4, 7$\pm$1) pc and a galactic motion of (U, V, W)=(-27$\pm$5, -3$\pm$5, 2$\pm$2) km s$^{-1}$,
using a right-handed coordinate system with X and U positive toward the galactic center.

In Figure 6,
we compare the galactic position (XYZ) and galactic velocity (UVW) of W2319+7645 (red star) with $1\sigma$ error bars
to the mean galactic position and mean galactic velocity of young moving groups 
with their $1\sigma$ dispersions. The $2\sigma$ (dashed line)
and $3\sigma$ (dotted line) dispersion values are shown for Argus.
The XYZ and UVW values for young moving groups are from \citet{Maloetal2013}.
W2319+7645 is consistent with the $\sim2\sigma$ dispersion values of the galactic position and galactic velocity of Argus
if we allow $3\sigma$ error bars for W2319+7645.
We use the BANYAN II web tool \citep{Maloetal2013,Gagneetal2014} to determine the probability 
of W2319+7645 belonging to one of the young moving groups. We used input parameters of
right ascension, declination, proper motion in right ascension, proper motion in declination, 
radial velocity, distance, error in proper motion in right ascension, error in proper motion in declination, 
error in radial velocity, error in distance, and include that W2319+7645 is younger than 1 Gyr.
The probability of membership for W2319+7645 to the young moving groups shown in Figure 6 is 79.44\% for 
membership in Argus and 20.56\% as a field star. W2319+7645 is a candidate for membership in Argus.

\begin{figure}
\caption{
The galactic position (XYZ) and galactic velocity (UVW) of W2319+7645 (red star) compared to 
the mean galactic position and mean galactic velocity for young moving groups 
from \citet{Maloetal2013}. The $1\sigma$ error bars are shown for W2319+7645 and the $1\sigma$
dispersion values are shown for the young moving groups. The $2\sigma$ (dashed line) 
and $3\sigma$ (dotted line) dispersion values are shown for Argus.
W2319+7645 is consistent with the $\sim2\sigma$ dispersion values of the galactic position and galactic velocity of Argus
if we allow $3\sigma$ error bars for W2319+7645.
The probability of W2319+7645 being a member of Argus is 79\% based on BANYAN II.
}
\begin{center}
\includegraphics[width=5.0in]{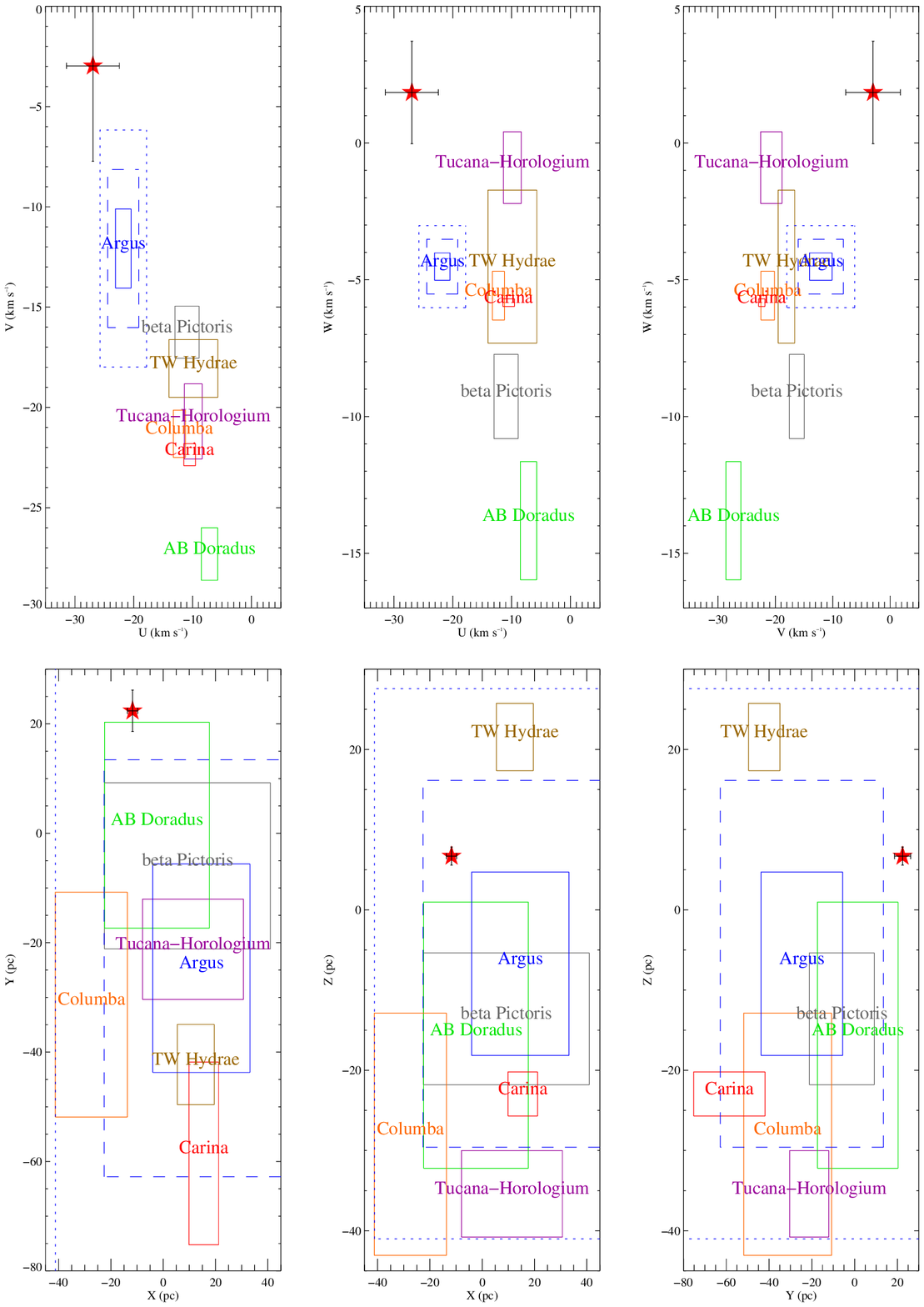}
\end{center}
\end{figure}

The spectral-type-effective-temperature relationship \citep{Looperetal2008a} gives a $T_{\rm eff}=1840\pm170$ K,
where the uncertainty in $T_{\rm eff}$ is twice the RMS in the spectral-type-effective-temperature
relation; we use double the RMS for the uncertainty in $T_{\rm eff}$ because W2319+7645 is not a
normal L dwarf. 
We use an IDL routine\footnote{Gagn\'e, Jonathan; Lafreni\`ere, David; Doyon, Ren\'e; Malo, Lison (2014): An IDL routine to estimate the mass of low-mass stars and brown dwarfs. figshare. \newline
http://dx.doi.org/10.6084/m9.figshare.899808 \newline
Retrieved 02:10, Oct 23, 2014 (GMT)} \citep{Gagneetal2014,Baraffeetal2003,Allardetal2013,Rajpurohitetal2013} to 
estimate the mass of W2319+7645. This routine uses
distance, an estimated age range, and 2MASS ($J$, $H$, $K_{\rm s}$) and WISE ($W1$, $W2$) photometry
to determine the most probable mass range in a likelihood analysis. Using our estimated distance, an 
age range of 30 to 50 Myr \citep{Gagneetal2014} based on membership in Argus, and the 2MASS and WISE photometry,
we find a mass of 12.1$\pm$0.4 M$_{\rm Jup}$ for W2319+7645, within the planetary mass regime.
Using an age estimate of $\sim100-120$ Myr for an intermediate-gravity L dwarf in the field, we find a 
mass of 21.9$\pm$2.0 M$_{\rm Jup}$ for W2319+7645, this firmly places W2319+7645 in the substellar regime. 
Table 1 gives the properties of W2319+7645.
\input{tab1.dat}

\section{CONCLUSIONS}
W2319+7645 is a young L dwarf of spectral type L4$\beta$, with a crude photometric distance 
estimate of $26.1\pm4.4$ pc.
An analysis of the 2MASS and WISE photometry of W2319+7645 reveals that the red WISE $W1-W2$ color 
provides additional evidence of youth while the 2MASS $J-K_{\rm s}$ color does not.
W2319+7645 is a candidate member of the young moving group Argus. Based on BANYAN II,
the space motion and position of W2319+7645 give a probability of 79\% membership in Argus and a probability
of 21\% as a field object. Based on membership in Argus, W2319+7645 has a mass estimate of 12.1$\pm$0.4 M$_{\rm Jup}$,
within the planetary mass regime.

Future work should include a parallax measurement, a more accurate radial velocity measurement, 
and near-infrared spectroscopy. A parallax measurement will provide a reliable distance estimate, since the 
photometric distance estimate is only a crude indicator for young L dwarfs.
A more accurate radial velocity 
measurement combined with a parallax measurement will help to provide a more accurate membership 
probability to the young moving group Argus.
Near-infrared spectroscopy will help to further characterize 
the youth of this object, following the near-infrared classification and low-gravity scheme of \citet{AllersLiu2013}.

\section{ACKNOWLEDGMENTS}
We thank the anonymous referee for a thorough report that significantly improved the manuscript.
This publication makes use of data products from the Wide-field Infrared Survey Explorer, which is a 
joint project of the University of California, Los Angeles, and the Jet Propulsion Laboratory/California 
Institute of Technology, and NEOWISE, which is a project of the Jet Propulsion Laboratory/California 
Institute of Technology. WISE and NEOWISE are funded by the National Aeronautics and Space Administration.
This publication makes use of data products from the Two Micron All Sky Survey, which is a joint project of the 
University of Massachusetts and the Infrared Processing and Analysis Center/California Institute of Technology, 
funded by the National Aeronautics and Space Administration and the National Science Foundation.
This research has made use of the NASA/ IPAC Infrared Science Archive, which is operated by the Jet Propulsion 
Laboratory, California Institute of Technology, under contract with the National Aeronautics and Space Administration.
This research has made use of the VizieR catalogue access tool, CDS, Strasbourg, France.

\bibliographystyle{apj}
\bibliography{/Users/Phil/research/bibliography}

\label{lastpage}

\end{document}

%% file: tab1.dat
\begin{deluxetable}{lc}
\tabletypesize{\scriptsize}
\tablecaption{Parameters of WISE J231921.92+764544.4}
\tablewidth{0pt}
\tablehead{
\colhead{Parameters} & \colhead{WISE J231921.92+764544.4}
}
\startdata
WISE R.A. (J2000) &23:19:21.93\\
WISE Decl. (J2000) &+76:45:44.5\\
WISE Epoch &2010.41\\
2MASS R.A. (J2000) &23:19:21.37\\
2MASS Decl. (J2000) &+76:45:43.8\\
2MASS Epoch &2000.46\\
2MASS $J$ (mag) &15.35$\pm$0.06\\
2MASS $H$ (mag) &14.39$\pm$0.05\\
2MASS $K_{\rm s}$ (mag) &13.65$\pm$0.05\\
WISE $W1$ (mag) &12.93$\pm$0.12\\
WISE $W2$ (mag) &12.45$\pm$0.02\\
WISE $W3$ (mag) &11.52$\pm$0.15\\
WISE $W4$ (mag) & $>$ 8.65\\
$J-H$ (mag) &0.97$\pm$0.08\\
$J-K_{\rm s}$ (mag) &1.71$\pm$0.08\\
$H-K_{\rm s}$ (mag) &0.74$\pm$0.07\\
$W1-W2$ (mag) &0.48$\pm$0.12\\
$W1-W3$ (mag) &1.41$\pm$0.19\\
$W2-W3$ (mag) &0.93$\pm$0.15\\
Spectral Type (Opt) & L4$\beta$\\
Distance (pc) &26.1$\pm$4.4\\
$\mu_{\alpha}$cos($\delta$) (mas yr$^{-1}$) &190$\pm$10\\
$\mu_{\delta}$ (mas yr$^{-1}$) &70$\pm$10\\
Total Apparent Motion ($\arcsec$ yr$^{-1}$) &0.20$\pm$0.01\\
${v}_{\rm tan}$ (km s$^{-1}$) &25$\pm$5\\
${v}_{\rm rad}$ (km s$^{-1}$) & 10$\pm$5\\
${v}_{\rm tot}$ (km s$^{-1}$) &27$\pm$7\\
$U$ (km s$^{-1}$) &-27$\pm$5\\
$V$ (km s$^{-1}$) &-3$\pm$5\\
$W$ (km s$^{-1}$) &2$\pm$2\\
$X$ (pc) &-12$\pm$2\\
$Y$ (pc) &22$\pm$4\\
$Z$ (pc) &7$\pm$1\\
$T_{\rm eff}$ (K) &1800$\pm$200\\
Age (Myr) &30$-$50\\
Mass (M$_{\rm Jup}$) &12.1$\pm$0.4\\
\enddata
\tablecomments{WISE data is from the AllWISE data release.}
\end{deluxetable}

%% file: msv7.bbl
\begin{thebibliography}{38}
\expandafter\ifx\csname natexlab\endcsname\relax\def\natexlab#1{#1}\fi

\bibitem[{{Allard} {et~al.}(2013){Allard}, {Homeier}, {Freytag},
  {Schaffenberger}, {}, \& {Rajpurohit}}]{Allardetal2013}
{Allard}, F., {Homeier}, D., {Freytag}, B., {Schaffenberger}, {}, W., \&
  {Rajpurohit}, A.~S. 2013, Memorie della Societa Astronomica Italiana
  Supplementi, 24, 128

\bibitem[{{Allers} \& {Liu}(2013)}]{AllersLiu2013}
{Allers}, K.~N., \& {Liu}, M.~C. 2013, ArXiv e-prints

\bibitem[{{Baraffe} {et~al.}(2003){Baraffe}, {Chabrier}, {Barman}, {Allard}, \&
  {Hauschildt}}]{Baraffeetal2003}
{Baraffe}, I., {Chabrier}, G., {Barman}, T.~S., {Allard}, F., \& {Hauschildt},
  P.~H. 2003, \aap, 402, 701

\bibitem[{{Basri}(2000)}]{Basri2000}
{Basri}, G. 2000, \araa, 38, 485

\bibitem[{{Burrows} {et~al.}(2001){Burrows}, {Hubbard}, {Lunine}, \&
  {Liebert}}]{Burrowsetal2001}
{Burrows}, A., {Hubbard}, W.~B., {Lunine}, J.~I., \& {Liebert}, J. 2001,
  Reviews of Modern Physics, 73, 719

\bibitem[{{Castro} \& {Gizis}(2012)}]{CastroGizis2012}
{Castro}, P.~J., \& {Gizis}, J.~E. 2012, \apj, 746, 3

\bibitem[{{Castro} {et~al.}(2013){Castro}, {Gizis}, {Harris}, {Mace},
  {Kirkpatrick}, {McLean}, {Pattarakijwanich}, \& {Skrutskie}}]{Castroetal2013}
{Castro}, P.~J., {Gizis}, J.~E., {Harris}, H.~C., {Mace}, G.~N., {Kirkpatrick},
  J.~D., {McLean}, I.~S., {Pattarakijwanich}, P., \& {Skrutskie}, M.~F. 2013,
  \apj, 776, 126

\bibitem[{{Cruz} {et~al.}(2009){Cruz}, {Kirkpatrick}, \&
  {Burgasser}}]{Cruzetal2009}
{Cruz}, K.~L., {Kirkpatrick}, J.~D., \& {Burgasser}, A.~J. 2009, \aj, 137, 3345

\bibitem[{{Cruz} {et~al.}(2007){Cruz}, {Reid}, {Kirkpatrick}, {Burgasser},
  {Liebert}, {Solomon}, {Schmidt}, {Allen}, {Hawley}, \&
  {Covey}}]{Cruzetal2007}
{Cruz}, K.~L., {et~al.} 2007, \aj, 133, 439

\bibitem[{{Delfosse} {et~al.}(1997){Delfosse}, {Tinney}, {Forveille},
  {Epchtein}, {Bertin}, {Borsenberger}, {Copet}, {de Batz}, {Fouque},
  {Kimeswenger}, {Le Bertre}, {Lacombe}, {Rouan}, \&
  {Tiphene}}]{Delfosseetal1997}
{Delfosse}, X., {et~al.} 1997, \aap, 327, L25

\bibitem[{{Dupuy} \& {Liu}(2012)}]{DupuyLiu2012}
{Dupuy}, T.~J., \& {Liu}, M.~C. 2012, ArXiv e-prints

\bibitem[{{Faherty} {et~al.}(2009){Faherty}, {Burgasser}, {Cruz}, {Shara},
  {Walter}, \& {Gelino}}]{Fahertyetal2009}
{Faherty}, J.~K., {Burgasser}, A.~J., {Cruz}, K.~L., {Shara}, M.~M., {Walter},
  F.~M., \& {Gelino}, C.~R. 2009, \aj, 137, 1

\bibitem[{{Faherty} {et~al.}(2013{\natexlab{a}}){Faherty}, {Cruz}, {Rice}, \&
  {Riedel}}]{Fahertyetal2013b}
{Faherty}, J.~K., {Cruz}, K.~L., {Rice}, E.~L., \& {Riedel}, A.
  2013{\natexlab{a}}, \memsai, 84, 955

\bibitem[{{Faherty} {et~al.}(2013{\natexlab{b}}){Faherty}, {Rice}, {Cruz},
  {Mamajek}, \& {N{\'u}{\~n}ez}}]{Fahertyetal2013}
{Faherty}, J.~K., {Rice}, E.~L., {Cruz}, K.~L., {Mamajek}, E.~E., \&
  {N{\'u}{\~n}ez}, A. 2013{\natexlab{b}}, \aj, 145, 2

\bibitem[{{Gagn{\'e}} {et~al.}(2014){Gagn{\'e}}, {Lafreni{\`e}re}, {Doyon},
  {Malo}, \& {Artigau}}]{Gagneetal2014}
{Gagn{\'e}}, J., {Lafreni{\`e}re}, D., {Doyon}, R., {Malo}, L., \& {Artigau},
  {\'E}. 2014, \apj, 783, 121

\bibitem[{{Gizis} {et~al.}(2011{\natexlab{a}}){Gizis}, {Burgasser}, {Faherty},
  {Castro}, \& {Shara}}]{Gizisetal2011a}
{Gizis}, J.~E., {Burgasser}, A.~J., {Faherty}, J.~K., {Castro}, P.~J., \&
  {Shara}, M.~M. 2011{\natexlab{a}}, \aj, 142, 171

\bibitem[{{Gizis} {et~al.}(2011{\natexlab{b}}){Gizis}, {Troup}, \&
  {Burgasser}}]{Gizisetal2011b}
{Gizis}, J.~E., {Troup}, N.~W., \& {Burgasser}, A.~J. 2011{\natexlab{b}},
  \apjl, 736, L34+

\bibitem[{{Gizis} {et~al.}(2012){Gizis}, {Faherty}, {Liu}, {Castro}, {Shaw},
  {Vrba}, {Harris}, {Aller}, \& {Deacon}}]{Gizisetal2012}
{Gizis}, J.~E., {et~al.} 2012, ArXiv e-prints

\bibitem[{{Hook} {et~al.}(2004){Hook}, {J{\o}rgensen}, {Allington-Smith},
  {Davies}, {Metcalfe}, {Murowinski}, \& {Crampton}}]{Hooketal2004}
{Hook}, I.~M., {J{\o}rgensen}, I., {Allington-Smith}, J.~R., {Davies}, R.~L.,
  {Metcalfe}, N., {Murowinski}, R.~G., \& {Crampton}, D. 2004, \pasp, 116, 425

\bibitem[{{Kirkpatrick}(2005)}]{Kirkpatrick2005}
{Kirkpatrick}, J.~D. 2005, \araa, 43, 195

\bibitem[{{Kirkpatrick} {et~al.}(2006){Kirkpatrick}, {Barman}, {Burgasser},
  {McGovern}, {McLean}, {Tinney}, \& {Lowrance}}]{Kirkpatricketal2006}
{Kirkpatrick}, J.~D., {Barman}, T.~S., {Burgasser}, A.~J., {McGovern}, M.~R.,
  {McLean}, I.~S., {Tinney}, C.~G., \& {Lowrance}, P.~J. 2006, \apj, 639, 1120

\bibitem[{{Kirkpatrick} {et~al.}(2001){Kirkpatrick}, {Dahn}, {Monet}, {Reid},
  {Gizis}, {Liebert}, \& {Burgasser}}]{Kirkpatricketal2001}
{Kirkpatrick}, J.~D., {Dahn}, C.~C., {Monet}, D.~G., {Reid}, I.~N., {Gizis},
  J.~E., {Liebert}, J., \& {Burgasser}, A.~J. 2001, \aj, 121, 3235

\bibitem[{{Kirkpatrick} {et~al.}(1999){Kirkpatrick}, {Reid}, {Liebert},
  {Cutri}, {Nelson}, {Beichman}, {Dahn}, {Monet}, {Gizis}, \&
  {Skrutskie}}]{Kirkpatricketal1999}
{Kirkpatrick}, J.~D., {et~al.} 1999, \apj, 519, 802

\bibitem[{{Kirkpatrick} {et~al.}(2000){Kirkpatrick}, {Reid}, {Liebert},
  {Gizis}, {Burgasser}, {Monet}, {Dahn}, {Nelson}, \&
  {Williams}}]{Kirkpatricketal2000}
---. 2000, \aj, 120, 447

\bibitem[{{Kirkpatrick} {et~al.}(2008){Kirkpatrick}, {Cruz}, {Barman},
  {Burgasser}, {Looper}, {Tinney}, {Gelino}, {Lowrance}, {Liebert},
  {Carpenter}, {Hillenbrand}, \& {Stauffer}}]{Kirkpatricketal2008}
---. 2008, \apj, 689, 1295

\bibitem[{{Kirkpatrick} {et~al.}(2011){Kirkpatrick}, {Cushing}, {Gelino},
  {Griffith}, {Skrutskie}, {Marsh}, {Wright}, {Mainzer}, {Eisenhardt},
  {McLean}, {Thompson}, {Bauer}, {Benford}, {Bridge}, {Lake}, {Petty},
  {Stanford}, {Tsai}, {Bailey}, {Beichman}, {Bloom}, {Bochanski}, {Burgasser},
  {Capak}, {Cruz}, {Hinz}, {Kartaltepe}, {Knox}, {Manohar}, {Masters},
  {Morales-Calder{\'o}n}, {Prato}, {Rodigas}, {Salvato}, {Schurr}, {Scoville},
  {Simcoe}, {Stapelfeldt}, {Stern}, {Stock}, \& {Vacca}}]{Kirkpatricketal2011}
---. 2011, \apjs, 197, 19

\bibitem[{{Looper} {et~al.}(2008){Looper}, {Gelino}, {Burgasser}, \&
  {Kirkpatrick}}]{Looperetal2008a}
{Looper}, D.~L., {Gelino}, C.~R., {Burgasser}, A.~J., \& {Kirkpatrick}, J.~D.
  2008, \apj, 685, 1183

\bibitem[{{Lucas} {et~al.}(2001){Lucas}, {Roche}, {Allard}, \&
  {Hauschildt}}]{Lucasetal2001}
{Lucas}, P.~W., {Roche}, P.~F., {Allard}, F., \& {Hauschildt}, P.~H. 2001,
  \mnras, 326, 695

\bibitem[{{Luhman}(2013)}]{Luhman2013}
{Luhman}, K.~L. 2013, \apjl, 767, L1

\bibitem[{{Luhman} {et~al.}(2012){Luhman}, {Loutrel}, {McCurdy}, {Mace},
  {Melso}, {Star}, {Young}, {Terrien}, {McLean}, {Kirkpatrick}, \&
  {Rhode}}]{Luhmanetal2012}
{Luhman}, K.~L., {et~al.} 2012, \apj, 760, 152

\bibitem[{{Malo} {et~al.}(2013){Malo}, {Doyon}, {Lafreni{\`e}re}, {Artigau},
  {Gagn{\'e}}, {Baron}, \& {Riedel}}]{Maloetal2013}
{Malo}, L., {Doyon}, R., {Lafreni{\`e}re}, D., {Artigau}, {\'E}., {Gagn{\'e}},
  J., {Baron}, F., \& {Riedel}, A. 2013, \apj, 762, 88

\bibitem[{{McGovern} {et~al.}(2004){McGovern}, {Kirkpatrick}, {McLean},
  {Burgasser}, {Prato}, \& {Lowrance}}]{McGovernetal2004}
{McGovern}, M.~R., {Kirkpatrick}, J.~D., {McLean}, I.~S., {Burgasser}, A.~J.,
  {Prato}, L., \& {Lowrance}, P.~J. 2004, \apj, 600, 1020

\bibitem[{{Rajpurohit} {et~al.}(2013){Rajpurohit}, {Reyl{\'e}}, {Allard},
  {Homeier}, {Schultheis}, {Bessell}, \& {Robin}}]{Rajpurohitetal2013}
{Rajpurohit}, A.~S., {Reyl{\'e}}, C., {Allard}, F., {Homeier}, D.,
  {Schultheis}, M., {Bessell}, M.~S., \& {Robin}, A.~C. 2013, \aap, 556, A15

\bibitem[{{Rebolo} {et~al.}(1992){Rebolo}, {Martin}, \&
  {Magazzu}}]{Reboloetal1992}
{Rebolo}, R., {Martin}, E.~L., \& {Magazzu}, A. 1992, \apjl, 389, L83

\bibitem[{{Rebolo} {et~al.}(1998){Rebolo}, {Zapatero Osorio}, {Madruga},
  {Bejar}, {Arribas}, \& {Licandro}}]{Reboloetal1998}
{Rebolo}, R., {Zapatero Osorio}, M.~R., {Madruga}, S., {Bejar}, V.~J.~S.,
  {Arribas}, S., \& {Licandro}, J. 1998, Science, 282, 1309

\bibitem[{{Reid} {et~al.}(2008){Reid}, {Cruz}, {Kirkpatrick}, {Allen},
  {Mungall}, {Liebert}, {Lowrance}, \& {Sweet}}]{Reidetal2008}
{Reid}, I.~N., {Cruz}, K.~L., {Kirkpatrick}, J.~D., {Allen}, P.~R., {Mungall},
  F., {Liebert}, J., {Lowrance}, P., \& {Sweet}, A. 2008, \aj, 136, 1290

\bibitem[{{Skrutskie} {et~al.}(2006){Skrutskie}, {Cutri}, {Stiening},
  {Weinberg}, {Schneider}, {Carpenter}, {Beichman}, {Capps}, {Chester},
  {Elias}, {Huchra}, {Liebert}, {Lonsdale}, {Monet}, {Price}, {Seitzer},
  {Jarrett}, {Kirkpatrick}, {Gizis}, {Howard}, {Evans}, {Fowler}, {Fullmer},
  {Hurt}, {Light}, {Kopan}, {Marsh}, {McCallon}, {Tam}, {Van Dyk}, \&
  {Wheelock}}]{Skrutskieetal2006}
{Skrutskie}, M.~F., {et~al.} 2006, \aj, 131, 1163

\bibitem[{{Wright} {et~al.}(2010){Wright}, {Eisenhardt}, {Mainzer}, {Ressler},
  {Cutri}, {Jarrett}, {Kirkpatrick}, {Padgett}, {McMillan}, {Skrutskie},
  {Stanford}, {Cohen}, {Walker}, {Mather}, {Leisawitz}, {Gautier}, {McLean},
  {Benford}, {Lonsdale}, {Blain}, {Mendez}, {Irace}, {Duval}, {Liu}, {Royer},
  {Heinrichsen}, {Howard}, {Shannon}, {Kendall}, {Walsh}, {Larsen}, {Cardon},
  {Schick}, {Schwalm}, {Abid}, {Fabinsky}, {Naes}, \& {Tsai}}]{Wrightetal2010}
{Wright}, E.~L., {et~al.} 2010, \aj, 140, 1868

\end{thebibliography}
